\documentclass[aps,prc,preprint,groupedaddress,amsmath,amssymb]{revtex4}

\usepackage{graphicx}
\usepackage{dcolumn}
\usepackage{bm}

\newcommand{\rmc}{\mathrm{c}}
\newcommand{\rmps}{\mathrm{p}}
\newcommand{\rmP}{\mathrm{P}}

\newcommand{\vtwo}{\hat{v}^{(2)}}
\newcommand{\vthree}{\hat{v}^{(3)}}

\begin{document}


\title{Study of the effect of the tensor correlation in oxygen isotopes with the charge- and parity-projected
Hartree-Fock method}


\author{Satoru Sugimoto}
\email[email: ]{satoru@ruby.scphys.kyoto-u.ac.jp} \affiliation{Kyoto
University, Kitashirakawa, Kyoto 606-8502, Japan}
\author{Kiyomi Ikeda}
\email[email: ]{k-ikeda@riken.jp}
\affiliation{The Institute of Physical and Chemical Research
(RIKEN), Wako, Saitama 351-0198, Japan}
\author{Hiroshi Toki}
\email[email: ]{toki@rcnp.osaka-u.ac.jp} \affiliation{Research
Center for Nuclear Physics (RCNP), Osaka University, Ibaraki, Osaka
567-0047, Japan}

\date{\today}

\begin{abstract}
Recently, we developed a mean-field-type framework which treats the
correlation induced by the tensor force. To exploit the tensor
correlation we introduce single-particle states with the parity and
charge mixing. To make a total wave function have a definite charge
number and a good parity, the charge number and parity projections
are performed. Taking a variation of the projected wave function
with respect to single-particle states a Hartree-Fock-like equation,
the charge- and parity-projected Hartree-Fock equation, is obtained.
In the charge- and parity-projected Hartree-Fock method, we solve
the equation selfconsistently. In this paper we extend the charge-
and parity-projected Hartree-Fock method to include a three-body
force, which is important to reproduce the saturation property of
nuclei in mean-field frameworks. We apply the charge- and
parity-projected Hartree-Fock method to sub-closed-shell oxygen
isotopes ($^{14}$O, $^{16}$O, $^{22}$O, $^{24}$O, and $^{28}$O) to
study the effect of the tenor correlation and its dependence on
neutron numbers. We obtain reasonable binding energies and matter
radii for these nuclei. It is found that relatively large energy
gains come from the tensor force in these isotopes and there is the
blocking effect by occupied neutron orbits on the tensor
correlation.
\end{abstract}


\maketitle

\section{Introduction}
The tensor force plays important roles in nuclear structure. The
study of nuclear matter using the Brueckner theory showed that the
tensor force is closely related to the binding mechanism and the
saturation property of nuclear matter \cite{bethe71}. The almost
exact calculations with very large model spaces exhibit a large
attractive energy comes from the tensor force
\cite{akaishi86,kamada01}. The tensor force is inferred to be
responsible for about a half of the single-particle spin-orbit
splitting in light nuclei \cite{terasawa60, ando81, myo05}.

Recently due to the development of experimental techniques, we can
access to various kinds of unstable nuclei experimentally. Those
experiments have been revealing that the shell structures of
unstable nuclei may change from those of stable nuclei
\cite{ozawa00,grawe04}. Considering the importance of the tensor
force in nuclear structure, the tensor force probably has an effect
on such structure changes of nuclei \cite{otsuka05}. Therefore, the
study of the effect of the tensor force in neutron-rich nuclei is
interesting and important.

To study the effect of the tensor force in a relatively large mass
region in the nuclear chart including unstable nuclei, we have
developed a framework based on a mean-field-type model
\cite{toki02,sugimoto04,ogawa04,ogawa06}. One of the most important
tensor correlations in closed-shell nuclei is a 2-particle--2-hole
(2p--2h) correlation. In a usual Hartree-Fock calculation, the
correlation induced by the tensor force can be treated restrictively
and the 2p--2h correlation is hard to be handled. The effect of the
tensor force is thought to be included in other kinds of forces like
the central, LS, and density-dependent forces in the usual
Hartree-Fock calculations. To treat the tensor force directly, we
introduce a single-particle state with the parity and charge mixing
considering the pseudoscalar and isovector characters of the pion,
which mediates the tensor force \cite{toki02, ogawa04}. Because a
total wave function made from such single-particle states with the
parity and charge mixing does not have a good parity and a definite
charge number, the parity and charge number projections are
performed before variation \cite{sugimoto04,ogawa06}. We call this
method the charge- and parity-projected Hartree-Fock (CPPHF) method.
In the previous studies we applied the CPPHF method to the alpha
particle and showed that the tensor correlation can be exploited in
the CPPHF method. The CPPHF method was also applied to $^8$Be to
study the effect of the tensor force on alpha clustering
\cite{sugimoto05}.

There are other attempts to treat the tensor correlation by
expanding usual model spaces like a mean-field model
\cite{akaishi04}, a shell model \cite{myo05}, the antisymmetrized
molecular dynamics (AMD) \cite{dote05}. Those studies including ours
showed the importance of the 2p--2h configuration mixing and
high-momentum components in single-particle states, which are not
treated in usual model space calculations. Neff and his
collaborators took a different approach to treat the tensor
correlation \cite{neff03}. They used the unitary correlation
operator method (UCOM) to make an effective interaction in a
moderate model space. In their effective interaction, the attractive
correlation by the tensor force is included in other forces like the
central and LS forces. They performed the calculations up to the
second order perturbation based on Hartree-Fock calculations using
their effective interaction and obtained a nice agreement of binding
energies over the whole mass region \cite{roth06}. Otsuka and his
collaborators \cite{otsuka05} showed that a particle-hole (p-h)
correlation by the tensor force is also important and changes
single-particle spin-orbit splittings of neutron (proton) orbits
with proton (neutron) numbers. The effect was inferred in old
Hartree-Fock calculations \cite{tarbutton68,bouyssy87}.

In the present paper, we apply the CPPHF method to sub-closed-shell
oxygen isotopes, $^{14}$O, $^{16}$O, $^{22}$O, $^{24}$O, and
$^{28}$O, which are assumed to have sub-closed-shell structures for
neutron orbits up to $0p_{3/2}$, $0p_{1/2}$, $0d_{5/2}$, $1s_{1/2}$,
and $0d_{3/2}$ respectively,  to see the dependence of the
correlation induced by the tensor force on neutron numbers. We
extend the CPPHF method to treat a three-body force, which is needed
to reproduce the saturation property of nuclei with relatively large
mass numbers. In Section~\ref{sec:cpphf} we explain the CPPHF method
with a three-body force. In Section~\ref{sec:result} the results of
the CPPHF method are presented. In Section~\ref{sec:summary} we
summarize the paper.
\section{Charge- and parity-projected Hartree-Fock
method with a three-body force}\label{sec:cpphf} In this section we
formulate the charge- and parity-projected Hartree-Fock (CPPHF)
method in the case where a three-body force exists. A Hamiltonian
for an A-body system with the two-body and three-body forces can be
written as
\begin{align}
 \hat{H} = \sum_{a=1}^A \hat{t}(x_a)
  +\sum_{a > b = 1}^A \vtwo (x_a,x_b)
  +\sum_{a>b>c=1}^A \vthree (x_a,x_b,x_c)
  \label{eq: hamiltonian}
  ,
\end{align}
where $\hat{t}$, $\vtwo$, and $\vthree$ are one-body, two-body, and
three-body operators respectively. $x$'s are coordinates including
spin and isospin. In the CPPHF method, we assume as single-particle
states the ones with the parity and charge mixing. It means each
single-particle wave function has both positive-parity and
negative-parity components and both proton and neutron components.
The single-particle wave function which consists of these four
components has the following form,
\begin{align}
 {\psi}_\alpha(x) =\sum_{p_\alpha=\pm}\ \sum_{t_{z\alpha}=\pm 1/2}
 \psi_{p_\alpha, t_{z\alpha}} (x).
  \label{eq: spwf}
\end{align}
In the above equation, $p_\alpha$ denotes parities, $+$ for positive
parity and $-$ for negative parity, and $t_{z\alpha}$ denotes
isospins, $+1/2$ for proton and $-1/2$ for neutron. In the CPPHF
method, we take as a wave function of an A-body system a Slater
determinant which consists of the single-particle states with the
parity and charge mixing,
\begin{align}
 \Psi^{\textrm{intr}}=\frac{1}{\sqrt{A !}}
  \hat{\mathcal{A}} \prod_{a=1}^{A}&
  \psi_{\alpha_a} (x_a).
  \label{eq: wf intr}
\end{align}
Here,  $\hat{\mathcal{A}}$ is the antisymmetrization operator.
Because $\Psi^{\textrm{intr}}$ does not have a good parity and a
definite charge number, we need to perform the projection operators
of parity ($\pm$) and charge number ($Z$) on $\Psi^{\textrm{intr}}$
to obtain the wave function with a good parity and a definite charge
number;
\begin{align}
 \Psi^{(\pm;Z)} = \hat{\mathcal{P}}^\mathrm{p}(\pm)
 \hat{\mathcal{P}}^\mathrm{c}(Z) \Psi^{\textrm{intr}}
 .
\end{align}
Here, $\hat{\mathcal{P}}^\mathrm{p}(\pm)$ is the parity-projection
operator, where $\hat{\mathcal{P}}^\mathrm{p}(+)$ projects out the
positive parity state and $\hat{\mathcal{P}}^\mathrm{p}(-)$ projects
out the negative parity one. $\hat{\mathcal{P}}^\mathrm{c}(Z)$ is
the charge-number-projection operator, which projects out the wave
function with a charge number $Z$. Therefore, $\Psi^{(\pm;Z)}$ has a
good parity ($\pm$) and a definite charge number ($Z$). The parity
projection operator $\hat{\mathcal{P}}^\mathrm{p}(\pm)$ is defined
as
\begin{align}
 \hat{\mathcal{P}}^\mathrm{p}(\pm) = \frac{1\pm \hat{P}}{2}
 \quad
  \left(\hat{P}=\prod_{a=1}^A \hat{p}_a
  \right)
  ,
 \label{eq: parityop}
\end{align}
where the total parity operator $\hat{P}$ is the product of the
parity operator  $\hat{p}_a$ for each single-particle state.
The charge projection operator $\hat{\mathcal{P}}^\mathrm{c}(Z)$ is
defined as
 \begin{align}
 \hat{\mathcal{P}}^\mathrm{c}(Z) = \frac{1}{2\pi} \int_0^{2\pi} d \theta
  e^{i(\hat{Z}-Z)\theta}
  = \frac{1}{2\pi} \int_0^{2\pi} d \theta
  e^{-i Z \theta}\hat{C}(\theta)
  \quad
  \left(
 \hat{Z} = \sum_{a=1}^A \frac{1+\tau^3_a}{2}
 \label{eq: chargeop}
    \right)
    ,
 \end{align}
where $\hat{Z}$ is the charge number operator, which is the sum of
the single-particle proton projection operator $(1+\tau_a^3)/2$, and
the charge-rotation operator is defined as
$\hat{C}(\theta)=e^{i\hat{Z}\theta}$.

We take the expectation value for the Hamiltonian $\hat{H}$ with the
projected wave function and obtain the energy functional,
\begin{align}
 E^{(\pm;Z)} =& \frac{\langle  \Psi^{(\pm;Z)}|\hat{H}
  |\Psi^{(\pm;Z)} \rangle}{\langle  \Psi^{(\pm;Z)} |\Psi^{(\pm;Z)}
 \rangle}
 = \frac{\langle  \Psi^\textrm{intr} |\hat{H}
  | \hat{\mathcal{P}}^\rmps(\pm) \hat{\mathcal{P}}^{\rmc}(Z) \Psi^\textrm{intr} \rangle}
 {\langle \Psi^\textrm{intr}|
  \hat{\mathcal{P}}^\rmps(\pm) \hat{\mathcal{P}}^{\rmc}(Z) \Psi^\textrm{intr}
 \rangle}
  \notag \\
 =& \frac{\frac{1}{4\pi}\int_0^{2 \pi} d \theta
  e^{-iZ\theta} \left(E^{(0)}(\theta)\pm E^{(\rmP)}(\theta)
  \right)}
 {\frac{1}{4\pi}\int_0^{2 \pi} d \theta
  e^{-iZ\theta} \left(n^{(0)}(\theta)\pm n^{(\rmP)}(\theta)
  \right)}.
 \label{eq: Etot}
\end{align}

The denominator in the right-hand side of the above equation is the
normalization of the total wave function,
\begin{align}
 n^{(\pm;Z)} \equiv \langle  \Psi^{(\pm;Z)} |\Psi^{(\pm;Z)} \rangle
 =\frac{1}{4\pi}\int_0^{2 \pi} d \theta
  e^{-iZ\theta} \left(n^{(0)}(\theta)\pm n^{(\rmP)}(\theta)
  \right).
\end{align}
Here, $n^{(0)}(\theta)$ is the determinant of the norm matrix
between the original single-particle wave functions
$\psi_{\alpha_a}$ and the charge-rotated single-particle wave
functions $\psi_{\alpha_a}(\theta)$. $n^{(\rmP)}(\theta)$ is the
determinant of the norm matrix between the original single-particle
wave functions $\psi_{\alpha_a}$ and the parity-inverted and
charge-rotated single-particle wave functions
$\psi_{\alpha_a}^{(\rmps)}(\theta)$.
\begin{align}
 n^{(0)}(\theta) &\equiv
 \langle \Psi^{\textrm{intr}}| \hat{C}(\theta)| \Psi^{\textrm{intr}} \rangle
 =
 \det B^{(0)}(\theta) \quad
 (B^{(0)}(\theta)_{ab} \equiv \langle
 \psi_{\alpha_a}|\psi_{\alpha_b}(\theta)\rangle)
 ,
 \notag \\
 n^{(\rmP)}(\theta) &\equiv
 \langle \Psi^{\textrm{intr}}|\hat{P}  \hat{C}(\theta) |\Psi^{\textrm{intr}} \rangle
 =
 \det B^{(\rmP)}(\theta) \quad
 (B^{(\rmP)}(\theta)_{ab}
 \equiv \langle \psi_{\alpha_a}|\psi_{\alpha_b}^{(\rmps)}(\theta)\rangle)
 .
\end{align}
The charge-rotated wave function $\psi_{\alpha_a}(x_b;\theta)$ and
the parity-inverted and charge-rotated wave function
$\psi_{\alpha_a}^{(\rmps)}(x_b;\theta)$ are defined as
\begin{align}
 \psi_{\alpha_a}(x_b;\theta) \equiv&
 e^{i\theta(1+\tau_b^3)/2}  \psi_{\alpha_a}(x_b)
\notag \\
 =& e^{i\theta}
 \psi_{p_\alpha=+, t_{z\alpha}=1/2}(x_b)
 +e^{i\theta}\psi_{p_\alpha=-, t_{z\alpha}=1/2}(x_b)
 \notag \\
 &+\psi_{p_\alpha=+, t_{z\alpha}=-1/2}(x_b)
 +\psi_{p_\alpha=-, t_{z\alpha}=-1/2}(x_b),
 \\
 \psi_{\alpha_a}^{(\rmps)}(x_b;\theta) \equiv&
 \hat{p}_b e^{i\theta(1+\tau_b^3)/2}  \psi_{\alpha_a}(x_b)
 \notag \\
  =& e^{i\theta}
 \psi_{p_\alpha=+, t_{z\alpha}=1/2}(x_b)
 -e^{i\theta}\psi_{p_\alpha=-, t_{z\alpha}=1/2}(x_b)
 \notag \\
 &+\psi_{p_\alpha=+, t_{z\alpha}=-1/2}(x_b)
 -\psi_{p_\alpha=-, t_{z\alpha}=-1/2}(x_b),
\end{align}
where $\hat{p}_b$ is the single-particle parity operator in
{(\ref{eq: parityop})} and $(1+\tau^3_b)/2$ is the single-particle
proton projection operator in (\ref{eq: chargeop}).

The numerator in the right-hand side of (\ref{eq: Etot}) is the
unnormalized total energy,
\begin{align}
 \langle  \Psi^{(\pm;Z)}|\hat{H}
  |\Psi^{(\pm;Z)} \rangle
 \equiv
 \frac{1}{4\pi}\int_0^{2 \pi} d \theta
  e^{-iZ\theta} \left(E^{(0)}(\theta)\pm E^{(\rmP)}(\theta)
  \right)
  \label{eq: unnormE}
  .
\end{align}
$E^{(0)}(\theta)$ in the right-hand side of (\ref{eq: unnormE}) has
a similar form as a simple Hartree-Fock energy but the
single-particle wave functions in the ket are modified by the charge
rotation,
\begin{align}
 &E^{(0)}(\theta)
  \equiv \langle  \Psi^{\textrm{intr}}|\hat{H} \hat{C}(\theta)
 |\Psi^{\textrm{intr}} \rangle
 \notag \\
 =&\sum_{a=1}^A
  \langle \psi_{\alpha_a} |\hat{t}|\tilde{\psi}_{\alpha_a}(\theta)\rangle
  +\sum_{a > b=1}^A
  \langle \psi_{\alpha_a}\psi_{\alpha_b}|\vtwo|
  \widehat{\tilde{\psi}_{\alpha_a}(\theta)\tilde{\psi}_{\alpha_b}(\theta)}
  \rangle
  \notag \\
  &+\sum_{a > b > c =1}^A
  \langle \psi_{\alpha_a}\psi_{\alpha_b}\psi_{\alpha_c}|\vthree|
  \widehat{\tilde{\psi}_{\alpha_a}(\theta)\tilde{\psi}_{\alpha_b}(\theta)\tilde{\psi}_{\alpha_c}(\theta)}
  \rangle
 .
\end{align}
Here, $\tilde{\psi}_{\alpha_a}(x;\theta)$ is the superposition of
$\psi_{\alpha_a}(x;\theta)$ weighted by the inverse of the
charge-rotated norm matrix $(B^{(0)}(\theta)^{-1})_{ba}$,
\begin{align}
 \tilde{\psi}_{\alpha_a}(x;\theta)=\sum_{b=1}^A
 \psi_{\alpha_b}(x;\theta)
 (B^{(0)}(\theta)^{-1})_{ba}.
 \label{eq: psitilde}
\end{align}
This summation for $\psi_{\alpha_b}(x;\theta)$ comes from the
antisymmetrization of the total wave function. The hats in the last
two terms represent the antisymmetrization as
\begin{align}
|\widehat{\tilde{\psi}_{\alpha_a}(\theta)\tilde{\psi}_{\alpha_b}(\theta)}\rangle
=& | \tilde{\psi}_{\alpha_a}(\theta)\tilde{\psi}_{\alpha_b}(\theta)
-\tilde{\psi}_{\alpha_b}(\theta)\tilde{\psi}_{\alpha_a}(\theta)
\rangle ,\\
|\widehat{\tilde{\psi}_{\alpha_a}(\theta)\tilde{\psi}_{\alpha_b}(\theta)\tilde{\psi}_{\alpha_c}(\theta)}\rangle
=& |
\tilde{\psi}_{\alpha_a}(\theta)\tilde{\psi}_{\alpha_b}(\theta)\tilde{\psi}_{\alpha_c}(\theta)
+\tilde{\psi}_{\alpha_b}(\theta)\tilde{\psi}_{\alpha_c}(\theta)\tilde{\psi}_{\alpha_a}(\theta)
\notag \\
&+\tilde{\psi}_{\alpha_c}(\theta)\tilde{\psi}_{\alpha_a}(\theta)\tilde{\psi}_{\alpha_b}(\theta)
-\tilde{\psi}_{\alpha_a}(\theta)\tilde{\psi}_{\alpha_c}(\theta)\tilde{\psi}_{\alpha_b}(\theta)
\notag \\
&-\tilde{\psi}_{\alpha_c}(\theta)\tilde{\psi}_{\alpha_b}(\theta)\tilde{\psi}_{\alpha_a}(\theta)
-\tilde{\psi}_{\alpha_b}(\theta)\tilde{\psi}_{\alpha_a}(\theta)\tilde{\psi}_{\alpha_c}(\theta)
\rangle .
\end{align}
$E^{(0)}(\theta=0)$ reduces to a simple Hartree-Fock energy.
$E^{(\rmP)}(\theta)$ in the right-hand side of (\ref{eq: unnormE})
has a similar form as $E^{(0)}(\theta)$ but
$\tilde{\psi}_{\alpha_a}(\theta)$'s are replaced by
$\tilde{\psi}_{\alpha_a}^{(\rmps)}(\theta)$'s,
\begin{align}
 &E^{(\rmP)}(\theta)
 \equiv
 \langle
 \Psi^{\textrm{intr}}|\hat{H} \hat{P} \hat{C}(\theta)|
 \Psi^{\textrm{intr}}\rangle
 \notag \\
 =&\sum_{a=1}^A
 \langle \psi_{\alpha_a} |\hat{t}|\tilde{\psi}_{\alpha_a}^{(\rmps)}(\theta)\rangle
 +\sum_{a > b=1}^A
 \langle \psi_{\alpha_a}\psi_{\alpha_b}|\hat{v}|
 \widehat{\tilde{\psi}_{\alpha_a}^{(\rmps)}(\theta)\tilde{\psi}_{\alpha_b}^{(\rmps)}(\theta)}
 \rangle
 \notag \\
 &+\sum_{a > b > c=1}^A\langle \psi_{\alpha_a}\psi_{\alpha_b}\psi_{\alpha_c}|\hat{v}|
 \widehat{\tilde{\psi}_{\alpha_a}^{(\rmps)}(\theta)\tilde{\psi}_{\alpha_b}^{(\rmps)}(\theta)
 \tilde{\psi}_{\alpha_c}^{(\rmps)}(\theta)}
 \rangle
 .
\end{align}
Here, $\tilde{\psi}_{\alpha_a}^{(\rmps)}(x;\theta)$ is the sum of
$\psi_{\alpha_a}^{(\rmps)}(x;\theta)$ weighted by the inverse of the
parity-inverted and charge-rotated norm matrix
$(B^{(\rmP)}(\theta)^{-1})_{ba}$,
\begin{align}
 \tilde{\psi}_{\alpha_a}^{(\rmps)}(x;\theta)=\sum_{b=1}^A
 \psi_{\alpha_b}^{(\rmps)}(x;\theta)
 (B^{(\rmP)}(\theta)^{-1})_{ba}.
\end{align}


We then take the variation of $E^{(\pm;Z)}$ with respect to a
single-particle wave function $\psi_{\alpha_a}$,
\begin{align}
 \frac{\delta}{\delta \psi_{\alpha_a}^\dagger (x_a)}
  \left\{E^{(\pm;Z)}-
 \sum_{b,c=1}^A \epsilon_{bc}
 \langle \psi_{\alpha_b} |\psi_{\alpha_c} \rangle
 \right\} = 0
 .
\end{align}
The Lagrange multiplier $\epsilon_{ab}$ is introduced to guarantee
the ortho-normalization of single-particle wave functions, $\langle
\psi_{\alpha_a}|\psi_{\alpha_b}\rangle =\delta_{\alpha_a,\alpha_b}$.
As the result, we obtain the following Hartree-Fock-like equation
with the charge and parity projections (the CPPHF equation) for each
$\psi_{\alpha_a}$,
\begin{align}
 \frac{1}{4\pi}\int_0^{2 \pi} d \theta e^{-iZ\theta}
 \Biggl[
 &
 n^{(0)}(\theta)
 \biggl\{
 \left( \hat{h}^{(1)}(x_a;\theta)+\hat{h}^{(2)}(x_a;\theta)+\hat{h}^{(3)}(x_a;\theta)
 \right) \tilde{\psi}_{\alpha_a}(x_a;\theta)
 \notag \\
 &
 -(E^{(\pm;Z)}-E^{(0)}(\theta))
 \tilde{\psi}_{\alpha_a}(x_a;\theta)
 -\sum_{b=1}^A\eta^{(0)}_{ba}(\theta) \tilde{\psi}_{\alpha_b}(x_a;\theta)
 \biggr\}
 \notag\\
 \pm&
 n^{(\rmP)}(\theta)
 \biggl\{\left(\hat{h}^{(1)(\rmps)}(x_a;\theta)+\hat{h}^{(2)(\rmps)}(x_a;\theta)+\hat{h}^{(3)(\rmps)}(x_a;\theta)
 \right)
 \notag \\
 &
 -(E^{(\pm;Z)}-E^{(\rmP)}(\theta))
 \tilde{\psi}_{\alpha_a}^{(\rmps)}(x_a;\theta)
 -\sum_{b=1}^A
 \eta^{(\rmP)}_{ba}(\theta)
 \tilde{\psi}_{\alpha_b}^{(\rmps)}(x_a;\theta)
 \biggr\}
 \Biggr]
 \notag\\
 &=n^{(\pm;Z)} \sum_{b=1}^A \epsilon_{ab} \psi_{\alpha_b}(x_a)
 ,
 \label{eq: cphf}
\end{align}
where $a=1,2,\dots,A$. Here, $\eta_{ab}^{(0)}(\theta)$ and
$\eta_{ab}^{(\rmP)}(\theta)$ are defined as follows,
  \begin{align}
  \eta_{ab}^{(0)}(\theta)
  \equiv &
  \langle \psi_{\alpha_a} |\hat{t}|\tilde{\psi}_{\alpha_b}(\theta)\rangle
  +\sum_{c=1}^A
  \langle \psi_{\alpha_a}\psi_{\alpha_c}|\vtwo|
  \widehat{\tilde{\psi}_{\alpha_b}(\theta)\tilde{\psi}_{\alpha_c}(\theta)}
  \rangle
  \notag \\
  &+\frac{1}{2}\sum_{c,d=1}^A
  \langle \psi_{\alpha_a}\psi_{\alpha_c}\psi_{\alpha_d}|\vthree|
  \widehat{\tilde{\psi}_{\alpha_b}(\theta)\tilde{\psi}_{\alpha_c}(\theta)\tilde{\psi}_{\alpha_d}(\theta)}
  \rangle
  ,
  \\
  \eta_{ab}^{(\rmP)}(\theta)
  \equiv &
  \langle \psi_{\alpha_a} |\hat{t}|\tilde{\psi}_{\alpha_b}^{(\rmps)}(\theta)\rangle
  +\sum_{c=1}^A
  \langle \psi_{\alpha_a}\psi_{\alpha_c}|\vtwo|
  \widehat{\tilde{\psi}_{\alpha_b}^{(\rmps)}(\theta)\tilde{\psi}_{\alpha_c}^{(\rmps)}(\theta)}
  \rangle
  \notag \\
  &+\frac{1}{2}\sum_{c,d=1}^A
  \langle \psi_{\alpha_a}\psi_{\alpha_c}\psi_{\alpha_d}|\vthree|
  \widehat{\tilde{\psi}_{\alpha_b}^{(\rmps)}(\theta)\tilde{\psi}_{\alpha_c}^{(\rmps)}(\theta)\tilde{\psi}_{\alpha_d}^{(\rmps)}(\theta)}
  \rangle
  .
  \end{align}
$\hat{h}^{(1)}$, $\hat{h}^{(2)}$, and $\hat{h}^{(3)}$ are the
single-particle operators originated from the one-body, two-body,
and three-body operators, which are defined as following,
\begin{align}
\hat{h}^{(1)} \tilde{\psi}_{\alpha_a}(x_a;\theta) \equiv&
\hat{t}(x_a) \tilde{\psi}_{\alpha_a}(x_a;\theta), \\
\hat{h}^{(2)} \tilde{\psi}_{\alpha_a}(x_a;\theta) \equiv&
\sum_{b=1}^A \Bigl\{ \langle \psi_{\alpha_b}|
\vtwo(x_{a})|\tilde{\psi}_{\alpha_b}(\theta) \rangle_1
\tilde{\psi}_{\alpha_a}(x_a;\theta) \notag \\
&-\langle \psi_{\alpha_b}|
\vtwo(x_{a})|\tilde{\psi}_{\alpha_a}(\theta) \rangle_1
\tilde{\psi}_{\alpha_b}(x_a;\theta)
\Bigr\}, \\
\hat{h}^{(3)} \tilde{\psi}_{\alpha_a}(x_a;\theta) \equiv&
\frac{1}{2} \sum_{b,c=1}^A \Bigl\{ \langle
\psi_{\alpha_b}\psi_{\alpha_c}|
\vthree(x_{a})|\widehat{\tilde{\psi}_{\alpha_b}(\theta)
\tilde{\psi}_{\alpha_c}(\theta)}\rangle_{1,2}
\tilde{\psi}_{\alpha_a}(x_a;\theta) \notag \\
&+\langle \psi_{\alpha_b}\psi_{\alpha_c}|
\vthree(x_{a})|\widehat{\tilde{\psi}_{\alpha_c}(\theta)
\tilde{\psi}_{\alpha_a}(\theta)}\rangle_{1,2}
\tilde{\psi}_{\alpha_b}(x_a;\theta) \notag \\
&+\langle \psi_{\alpha_b}\psi_{\alpha_c}|
\vthree(x_{a})|\widehat{\tilde{\psi}_{\alpha_a}(\theta)
\tilde{\psi}_{\alpha_b}(\theta)}\rangle_{1,2}
\tilde{\psi}_{\alpha_c}(x_a;\theta)\Big\} .
\end{align}
The expressions of $\hat{h}^{(1)(\rmps)}$, $\hat{h}^{(2)(\rmps)}$,
and $\hat{h}^{(3)(\rmps)}$ are obtained from those of
$\hat{h}^{(1)}$, $\hat{h}^{(2)}$, and $\hat{h}^{(3)}$ by replacing
$\tilde{\psi}_\alpha (x,\theta)$ with $\tilde{\psi}^{(\rmps)}_\alpha
(x,\theta)$. The notations for the integration of the two-body
matrix elements,
\begin{align}
 \langle \psi_{\alpha_b}|\vtwo(x_a)|\psi_{\alpha_c} \rangle_{1}
  =\int dx_1
  \psi^\dagger_{\alpha_b}(x_1) \vtwo (x_a,x_1)
  \psi_{\alpha_c}(x_1),
\end{align}
and for that of the three-body matrix elements,
\begin{align}
 &\langle \psi_{\alpha_b}\psi_{\alpha_c}|\vthree(x_a)|\psi_{\alpha_d}\psi_{\alpha_e} \rangle_{1,2}
  \notag \\
  &=\int dx_1 \int dx_2
  \psi^\dagger_{\alpha_b}(x_1)\psi^\dagger_{\alpha_c}(x_2)
  \vthree (x_a,x_1,x_2)
  \psi_{\alpha_d}(x_1) \psi_{\alpha_e}(x_2)
\end{align}
are introduced. The system of the coupled equations (\ref{eq: cphf})
for $a=1,\cdots,A$ are solved selfconsistently.

We give here the expressions for the expectation value of the
kinetic energy $\langle \hat{T} \rangle^{(\pm;Z)}$ with the center
of mass correction, that of the two-body potential energy $\langle
\vtwo \rangle^{(\pm;Z)}$, and that of three-body potential energy
$\langle \vthree \rangle^{(\pm;Z)}$ for the later convenience.
\begin{align}
 &\langle \hat{T} \rangle^{(\pm;Z)}
 =
 \frac{1}{4 \pi n^{(\pm;Z)}} \int_0^{2 \pi} d \theta e^{-i Z \theta}
 \notag \\
 \times \Biggl[&n^{(0)}(\theta)\Biggl\{
 \sum_{a=1}^A
 \langle \psi_{\alpha_a} |\hat{t}|\tilde{\psi}_{\alpha_a}(\theta)\rangle
 +\sum_{a > b=1}^A
 \langle \psi_{\alpha_a}\psi_{\alpha_b}|\frac{\hbar^2}{A M}
 \boldsymbol{\nabla}_a\cdot\boldsymbol{\nabla}_b|
 \widehat{\tilde{\psi}_{\alpha_a}(\theta)\tilde{\psi}_{\alpha_b}(\theta)}
 \rangle
 \Biggr\}
 \notag\\
 \pm&
 n^{(\rmP)}(\theta)
 \Biggl\{\sum_{a=1}^A
 \langle \psi_{\alpha_a}
 |\hat{t}|\tilde{\psi}_{\alpha_a}^{(\rmps)}(\theta)\rangle
 +\sum_{a > b=1}^A
 \langle \psi_{\alpha_a}\psi_{\alpha_b}|\frac{\hbar^2}{A M}
 \boldsymbol{\nabla}_a\cdot\boldsymbol{\nabla}_b|
 \widehat{\tilde{\psi}_{\alpha_a}^{(\rmps)}(\theta)\tilde{\psi}_{\alpha_b}^{(\rmps)}(\theta)}
 \rangle
 \Biggr\}
 \Biggr],
\end{align}
\begin{align}
 \langle \vtwo \rangle^{(\pm;Z)}
 =&
 \frac{1}{4 \pi n^{(\pm;Z)}} \int_0^{2 \pi} d \theta e^{-i Z \theta}
 \sum_{a > b=1}^A\Biggl\{
 n^{(0)}(\theta)
 \langle \psi_{\alpha_a}\psi_{\alpha_b}|\vtwo|
 \widehat{\tilde{\psi}_{\alpha_a}(\theta)\tilde{\psi}_{\alpha_b}(\theta)}
 \rangle
 \notag \\
 &\pm
 n^{(\rmP)}(\theta)
 \langle \psi_{\alpha_a}\psi_{\alpha_b}|\vtwo|
 \widehat{\tilde{\psi}_{\alpha_a}^{(\rmps)}(\theta)\tilde{\psi}_{\alpha_b}^{(\rmps)}(\theta)}
 \rangle
 \Biggr\},
\end{align}
\begin{align}
 \langle \vthree \rangle^{(\pm;Z)}
 =&
 \frac{1}{4 \pi n^{(\pm;Z)}} \int_0^{2 \pi} d \theta e^{-i Z \theta}
 \sum_{a > b > c=1}^A\Biggl\{
 n^{(0)}(\theta)
 \langle \psi_{\alpha_a}\psi_{\alpha_b}\psi_{\alpha_c}|\vthree|
 \widehat{\tilde{\psi}_{\alpha_a}(\theta)\tilde{\psi}_{\alpha_b}(\theta)\tilde{\psi}_{\alpha_c}(\theta)}
 \rangle
 \notag \\
 &\pm
 n^{(\rmP)}(\theta)
 \langle \psi_{\alpha_a}\psi_{\alpha_b}\psi_{\alpha_c}|\vthree|
 \widehat{\tilde{\psi}_{\alpha_a}^{(\rmps)}(\theta)\tilde{\psi}_{\alpha_b}^{(\rmps)}(\theta)\tilde{\psi}_{\alpha_c}^{(\rmps)}(\theta)}
 \rangle
 \Biggr\}.
\end{align}

\section{Applications to sub-closed-shell oxygen isotopes}\label{sec:result}
In this section, we apply the charge- and parity-projected
Hartree-Fock (CPPHF) method formulated in the last section to the
ground states of the sub-closed oxygen isotopes, $^{14}$O, $^{16}$O,
$^{22}$O, $^{24}$O and $^{28}$O. We assume the parities of these
nuclei in the ground states are positive. Those nuclei are assumed
to have the closed shell up to $0p_{1/2}$ for proton and sub-closed
or closed shells up to $0p_{3/2}$ ($^{14}$O), $0p_{1/2}$ ($^{16}$O),
$0d_{5/2}$ ($^{22}$O), $1s_{1/2}$ ($^{24}$O) and $0d_{3/2}$
($^{28}$O) for neutron. Although $^{28}$O is know to be unbound from
the experiment, we calculate the nucleus to study the
shell-configuration dependence of the contribution from the tensor
force theoretically. We assume the spherical symmetry. In this case,
only the total angular momentum $j$ is a good quantum number of a
single-particle state, because parities and charges are mixed in
intrinsic single-particle states. An intrinsic wave function can be
written in the following form,
\begin{align}
 \Psi^{\textrm{intr}} = \hat{\mathcal{A}} \prod_{0\leq j \leq j^\text{max}} \prod_{-j\leq m \leq j} \prod_{1 \leq n_j \leq n_j^\text{max}}
 \psi_{n_j j m}(x).
 \label{eq: wf intr alpha}
\end{align}
A single-particle wave function $\psi_{n_j j m}$ is composed of four
components, proton and positive parity, proton and negative parity,
neutron and positive parity, and neutron and negative parity,
\begin{align}
 {\psi_{n_j j m}}(x) =
  \sum_{t_z=\pm \frac{1}{2}}\bigl(
             &{
  \phi_{n_j j l_+ t_z}(r) \mathcal{Y}_{j l_+ m}(\Omega)\zeta(t_z)}
  +{
  \phi_{n_j j l_- t_z}(r) \mathcal{Y}_{j l_- m}(\Omega)
  \zeta(t_z)}
  \bigr)
  .
 \label{eq:spwf}
\end{align}
Here, $\mathcal{Y}_{jlm}(\Omega)$ is the eigenfunction of the total
angular momentum $\boldsymbol{j}=\boldsymbol{l}+\boldsymbol{s}$,
$\zeta(t_z)$ is the isospin wave function with $t_z=1/2$ for proton
and $t_z=-1/2$ for neutron. $l_+$ and $l_-$ are the orbital angular
momentum with positive parity and negative parity respectively. For
example $l_+=0$ and $l_-=1$ for $j=1/2$, $l_+=2$ and $l_-=1$ for
$j=3/2$, and so on. Eq.~(\ref{eq:spwf}) indicates that in the
present calculation assuming the spherical symmetry only the
correlations among the same $j$ orbits can be treated. It is a
limitation of the CPPHF method with the spherical symmetry. For the
calculation of $^{16}$O four single-particle states with $j=1/2$ and
two states with $j=3/2$ are included. They correspond to $\pi
s_{1/2}$, $\nu s_{1/2}$, $\pi p_{1/2}$ and  $\nu p_{1/2}$ for
$j=1/2$ and $\pi p_{3/2}$ and  $\nu p_{3/2}$ for $j=3/2$. Because
parities and charges are mixed in single-particle states such a
classification is approximately valid. For the calculation of
$^{14}$O one state with $j_{1/2}$ is subtracted and for the
calculation of $^{22}$O, $^{24}$O and $^{28}$O new states with
$j=5/2$, $j=1/2$ and $j=3/2$ are added one by one.

In the present study we use the modified Volkov force No.~1 (MV1
force) \cite{ando80} for the central potential and the G3RS force
\cite{tamagaki68} for the non-central forces. The MV1 force is the
modified version of the Volkov force No.~1 \cite{volkov65} and
includes the $\delta$-function-type three-body force,
\begin{align}
\hat{v}^{(3)}(x_a,x_b,x_c)=t_3
\delta(\boldsymbol{x}_a-\boldsymbol{x}_b)
\delta(\boldsymbol{x}_b-\boldsymbol{x}_c) . \label{eq: delta3B}
\end{align}
The Majorana parameter in the MV1 force is fixed to 0.6. The G3RS is
determined from the nucleon-nucleon scattering data. The effect of
the tensor force is effectively included in the MV1 force because
the MV1 force is determined so as to reproduce the binding energy of
$^{16}$O in the absence of the tensor force. The effect of the
tensor force is thought to appear in the $^3$E channel of the
central force as attraction. Therefore we multiply the attraction
part in the $^3$E channel of the central force by $x_\text{C}$. We
also multiply the three-body-force part by $x_\text{3B}$. The
$\delta$-function-type three-body force in Eq.~(\ref{eq: delta3B})
reduces to the density-dependent two-body force,
$\frac{1}{6} t_3 \rho (\frac{\boldsymbol{x}_a+\boldsymbol{x}_b}{2})
\frac{1+P_\sigma(ab)}{2} \delta
(\boldsymbol{x}_a-\boldsymbol{x}_b)$, for the wave functions of
even-even nuclei with time-reversal symmetry in the Hartree-Fock
level \cite{vautherin72}. $P_\sigma$ is the spin-exchange operator
and $\rho$ is a single-particle density. In this case the
density-dependent force only acts on the $^3$E channel. In the MV1
force $t_3$ is positive and the density-dependent force has the
repulsive effect on the $^3$E channel. The LS forces determined from
the NN scattering data is usually weak to be used in the mean filed
(Hartree-Fock) calculation. Hence, we multiply the LS force by 2. In
this case the strength of the LS force is comparable to those
adopted in the Skyrme forces and the Gogny forces \cite{patyk99}.

We also multiply the $\tau_1\cdot\tau_2$ part of the tensor force a
numerical factor $x_\text{T}$ as in the previous study
\cite{sugimoto04}, because the CPPHF method is a mean-field-type
calculation and can only take into account the correlations induced
by limited couplings among single-particle states. Actually, in the
spherical symmetry the 2-particle--2-hole (2p--2h) correlation which
can be treated in the CPPHF method is like
($j_{p_1}j_{p_2}j_{h_1}^{-1}j_{h_2}^{-1}$) with $j_{p_1}=j_{h_1}$
and $j_{p_2}=j_{h_2}$. However 2p-2h configurations with
$j_{p_1}\neq j_{h_1}$ and $j_{p_2}\neq j_{h_2}$ are also important
\cite{myo05,myo06}. Furthermore, other effects may enhance the
tensor correlation as mentioned in our previous paper
\cite{sugimoto04}. To take into account such effects effectively, we
take $x_\text{T}$=1.5 (the strong tensor force case) in addition to
$x_\text{T}$=1.0 (the normal tensor force case). $x_\text{C}$ and
$x_\text{3B}$ are determined to reproduce the binding energy and the
charge radius of $^{16}$O for each $x_\text{T}$.

We expand single-particle wave functions in the Gaussian basis as in
the previous study \cite{sugimoto04}. The number of the Gaussian
basis used is 10 for each orbit with the minimum range 0.5 fm and
the maximum range 10 fm. The CPPHF equation (\ref{eq: cphf}) is
solved by the gradient or the damped gradient method \cite{CNP1}.
The convergence of the calculation is quite slow for the case with a
large difference between a proton number ($Z$) and a neutron number
($N$). as in $^{28}$O. To remedy it, the quadratic constraint
potential term \cite{Flocard73} for $Z$,
\begin{align}
\langle\Psi^{\text{intr}}|\frac{\lambda}{2} (\hat{Z}-Z)^2
|\Psi^{\text{intr}}\rangle,
\end{align}
is added to the energy functional (\ref{eq: Etot}). The addition of
the constrained potential makes the convergence faster. The value of
$\lambda$ is taken as 1000 MeV.

\subsection{Results for $^{16}$O}
We first take $^{16}$O as a typical example and show the effect of
the tensor force in the CPPHF method.
\begin{table}
\caption{\label{table: 16O} Results for $^{16}$O in the HF and CPPHF
method. $x_\text{T}$ is a numerical factor multiplied to the
$\tau_1\cdot\tau_2$ part of the tensor force. $E$ and $T$ are the
total energy and the total kinetic energy respectively.
$V_\text{C}$, $V_\text{3B}$, $V_\text{T}$, $V_\text{LS}$ and
$V_\text{Coul}$ are the potential energies form the central, the
three-body force, the tensor force, the LS force, and the Coulomb
force respectively. Those are give in the unit of MeV. $R_\text{c}$
is the root-mean-square charge radius in the unit of fm. \label{tbl:
16O}}
\begin{ruledtabular}
\begin{tabular}{cccccccc}
 &{$x_\text{T}$} &{$E$} & {$T$} &
{$V_\text{C}+V_\text{3B}+V_\text{Coul}$} & {$V_\text{T}$}
&{$V_\text{LS}$} &{$R_\text{c}$}
\\\hline
HF & 1.0  & -124.1  & 230.0  & -353.2  & 0.0  & -0.9  & 2.73
\\ CPPHF & 1.0  & -127.1  & 237.1  & -351.6  & -11.7  & -1.0
& 2.73  \\
CPPHF & 1.5 & -127.6  & 253.9  & -342.2  & -38.3  & -1.0 & 2.73
\end{tabular}
\end{ruledtabular}
\end{table}
In Table~\ref{tbl: 16O} the results for $^{16}$O in the Hartree-Fock
(HF) and the charge- and parity-projected Hartree-Fock (CPPHF)
schemes are shown. In the HF scheme we fix $x_\text{T}$,
$x_\text{C}$, and $x_\text{3B}$ to 1.0. In the CPPHF scheme
$x_\text{C}$ and $x_\text{3B}$ are 1.025 and 1.25 for the normal
tensor force ($x_\text{T} =1.0$) case, and 1.040 and 1.55 for the
strong tensor force ($x_\text{T} = 1.5$) case. The root-mean-square
charge radius $R_\rmc$ is calculated from the proton
root-mean-square radius $R_\rmps$ as $R_\rmc =
\sqrt{R_\rmps^2+0.64}$. This approximation for $R_\rmc$ corresponds
to assume the charge radius of proton as 0.80 fm. In the HF
calculation the expectation value for the potential energy from the
tensor force $V_\text{T}$ is negligibly small. If we perform the
charge and parity projection before variation (the CPPHF case),
$V_\text{T}$ comes out to be a sizable value. It becomes about 10
MeV for the normal tensor force case and about 40 MeV for the strong
tensor force case. This result indicates that the CPPHF method is
effective to treat the correlation from the tensor force. In the
CPPHF cases the kinetic energy $T$ becomes larger than in the HF
case. In the CPPHF scheme, to gain the tensor correlation energy the
opposite-parity components compared to the simple shell-model
picture have to be mixed into single-particle states. This mixing
causes an over shell correlation and, as the result, the kinetic
energy becomes larger. A similar tendency is also observed in the
alpha particle case \cite{sugimoto04, ogawa06}.

\begin{figure}
\includegraphics{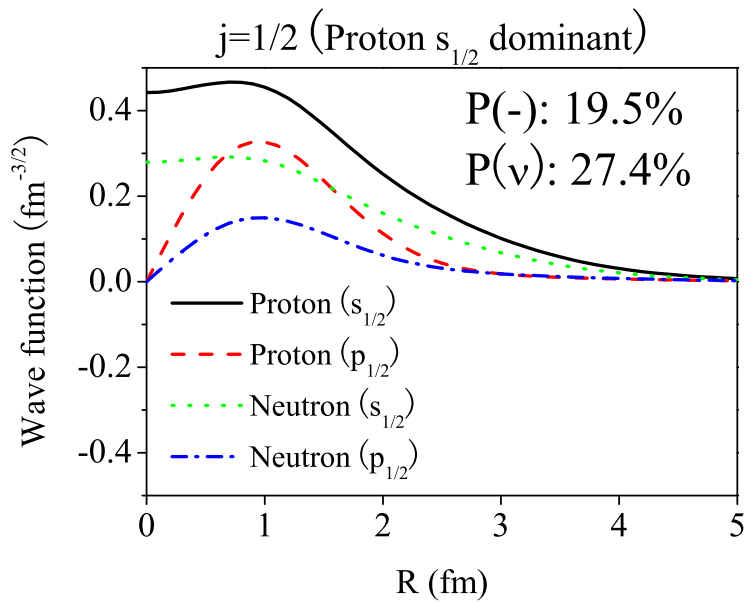}
\caption{\label{fig:o16wfs1ht15} (Color online) Intrinsic
single-particle wave function of the $j=1/2$ state with an $s_{1/2}$
proton component as a dominant one in $^{16}$O as a function of the
radial distance $R$. The solid, dashed, dotted and dashed-dotted
curves correspond $s_{1/2}$ proton, $p_{1/2}$ proton, $s_{1/2}$
neutron, and $p_{1/2}$ neutron respectively. $P(-)$ and $P(\nu)$ are
the mixing probabilities of the negative-party and the neutron
components respectively.}
\end{figure}
\begin{figure}
\includegraphics{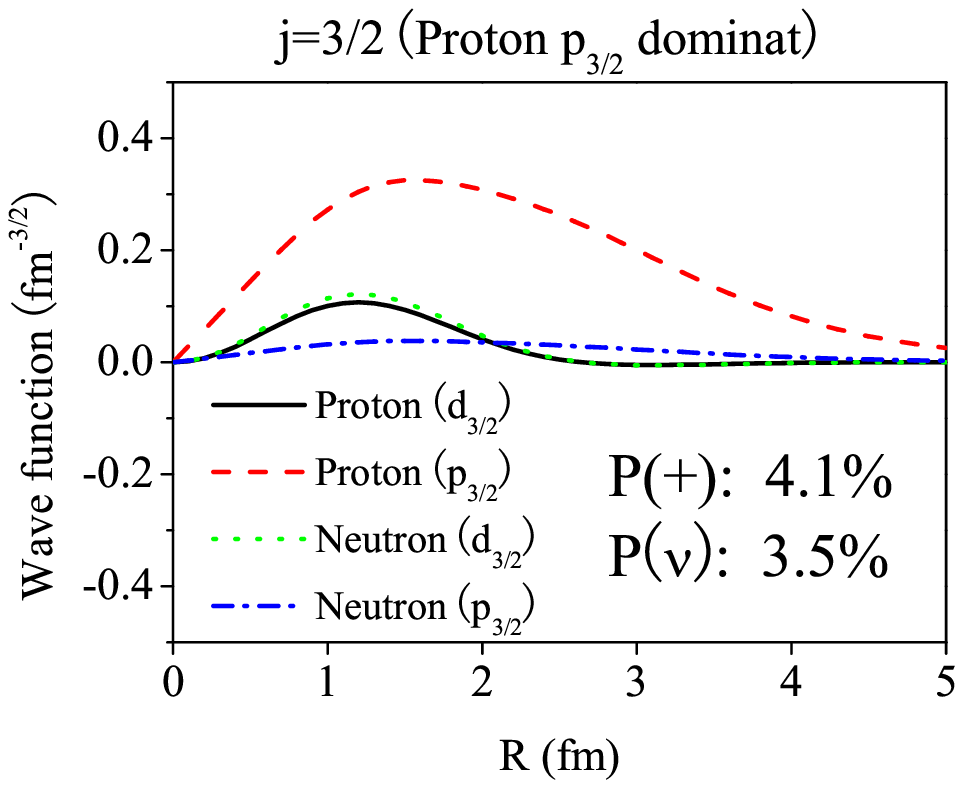}
\caption{\label{fig:o16wfp3ht15} (Color online) Intrinsic
single-particle wave function of the $j=3/2$ state with a $p_{3/2}$
proton component as a dominant one in $^{16}$O as a function of the
radial distance $R$. The solid, dashed, dotted and dashed-dotted
curves correspond $d_{3/2}$ proton, $p_{3/2}$ proton, $d_{3/2}$
neutron, and $p_{3/2}$ neutron respectively. $P(+)$ and $P(\nu)$ are
the mixing probabilities of the positive-party and the neutron
components respectively.}
\end{figure}
\begin{figure}
\includegraphics{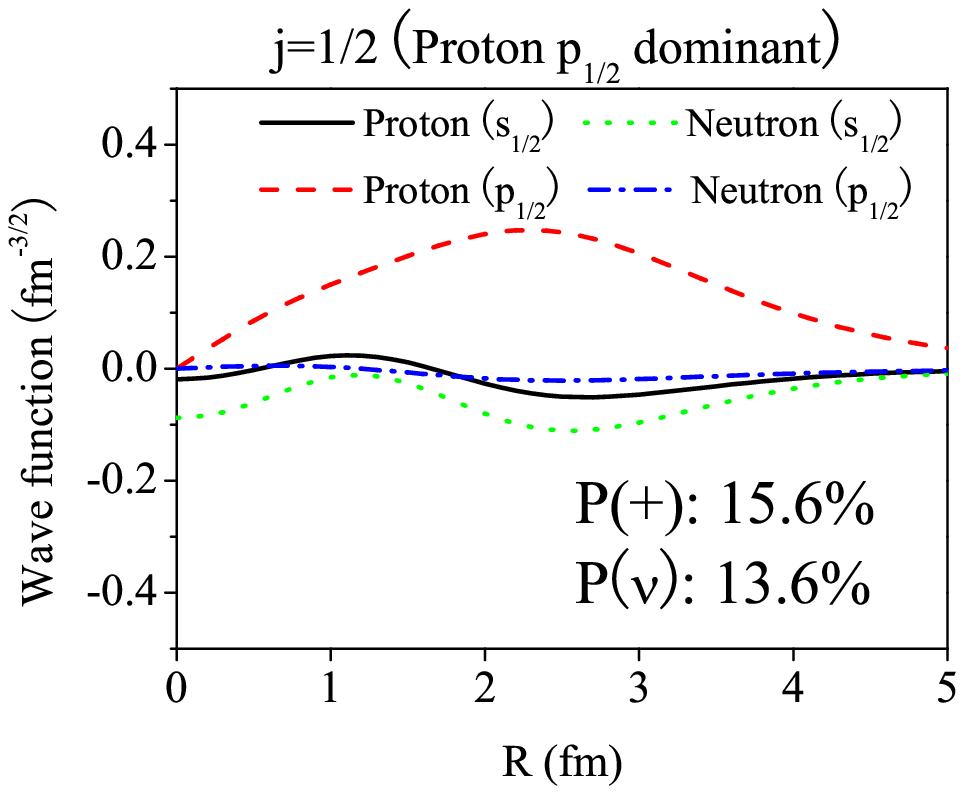}
\caption{\label{fig:o16wfp1ht15} (Color online) Intrinsic
single-particle wave function of the $j=1/2$ state with a $p_{1/2}$
proton component as a dominant one in $^{16}$O as a function of the
radial distance $R$. The solid, dashed, dotted and dashed-dotted
curves correspond $s_{1/2}$ proton, $p_{1/2}$ proton, $s_{1/2}$
neutron, and $p_{1/2}$ neutron respectively. $P(+)$ and $P(\nu)$ are
the mixing probabilities of the positive-party and the neutron
components respectively.}
\end{figure}
In Figs.~\ref{fig:o16wfs1ht15}, \ref{fig:o16wfp3ht15} and
\ref{fig:o16wfp1ht15} the intrinsic single-particle wave functions
in $^{16}$O with the strong tensor force ($x_\text{T}=1.5$) are
plotted. The wave functions plotted have proton components as
dominant ones. The wave function in Fig.~\ref{fig:o16wfs1ht15} has a
proton $s_{1/2}$ component as a dominant component. The mixing
probabilities for negative parity ($p_{1/2}$) and neutron are 19.5\%
and 27.4\% respectively. From the figure you can see that the spread
of the $p_{1/2}$ components are smaller than the $s_{1/2}$ ones. It
indicates that the opposite-parity components induced by the tensor
force have high-momentum components. The wave function which has a
proton $p_{3/2}$ component as a dominant one is plotted in
Fig.~\ref{fig:o16wfp3ht15} and the shrinkage of $d_{3/2}$ components
compared to $p_{3/2}$ ones are clearly seen also. In the wave
function which has a proton $p_{1/2}$ component as a dominant one,
which is plotted in Fig.~\ref{fig:o16wfp1ht15}, the shrinkage is not
so clear compared to the previous two cases, probably because of the
orthogonality condition to the wave function to the first $j=1/2$
state in Fig.~\ref{fig:o16wfs1ht15}.

In the alpha particle case, $p_{1/2}$ components mixing into
$s_{1/2}$ ones are also compact in size \cite{akaishi04, sugimoto04,
ogawa06}. The importance of this shrinkage is confirmed in a shell
model calculation \cite{myo05} and the antisymmetrized molecular
dynamics (AMD) calculation \cite{dote05}, too. The present result
infer that the shrinkage of mixing single-particle wave functions is
generally important in heavier-mass region. There are also wave
functions with a neutron component as a dominant one. The general
tendency just mentioned above is almost the same if proton and
neutron are interchanged.

\begin{figure}
\includegraphics{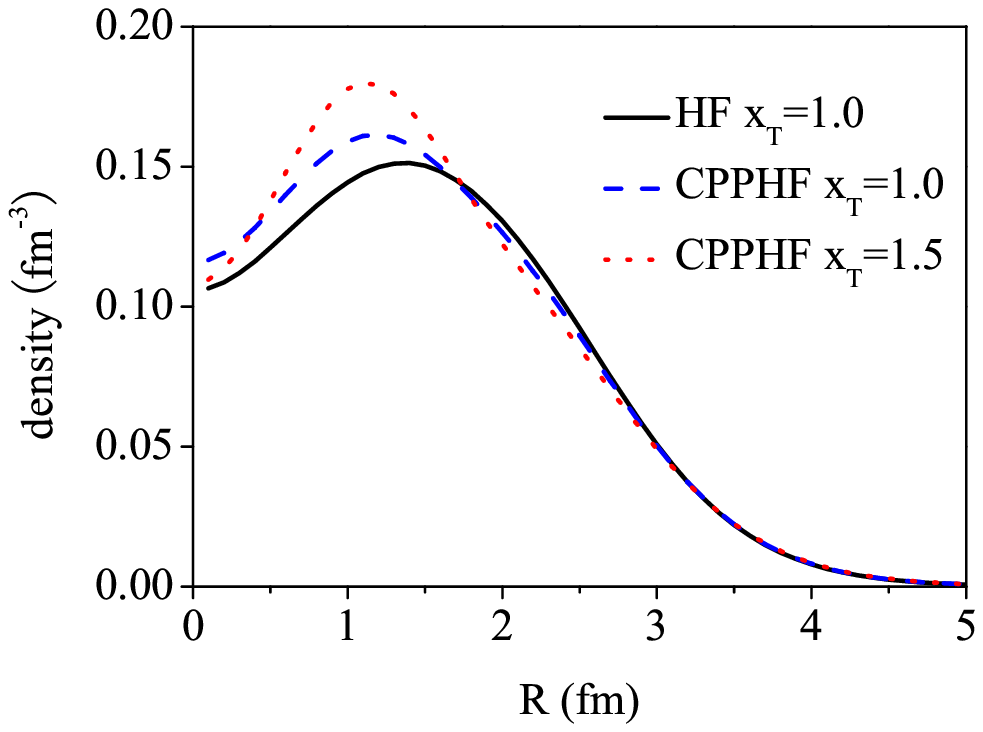}
\caption{\label{fig:o16density} (Color online) Densities for
$^{16}$O in the Hartree-Fock (HF) and the charge- and
parity-projected Hartree-Fock (CPPHF) methods as a function of the
radial distance ($R$). The solid, the dashed, and the dotted curves
correspond to the HF calculation, the CPPHF calculation with the
normal tensor force ($x_\text{T}=1.0$), and the CPPHF calculation
with the strong tensor force ($x_\text{T}=1.5$).}
\end{figure}
\begin{figure}
\includegraphics{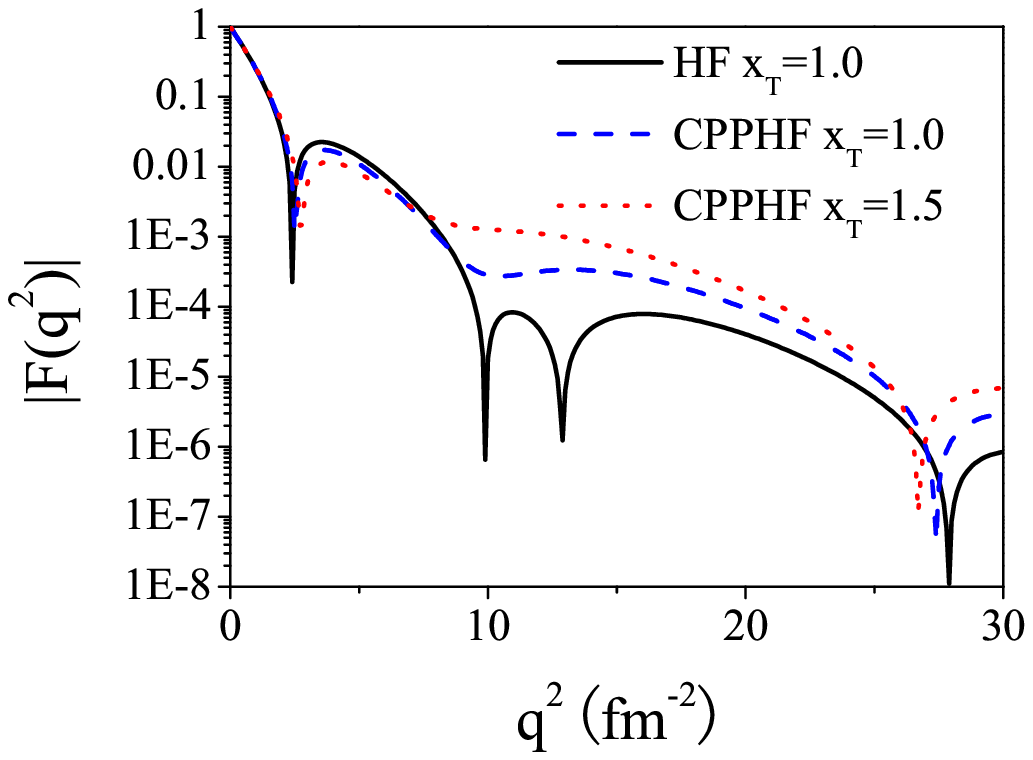}
\caption{\label{fig:o16CFF} (Color online) Absolute values of the
charge form factor for $^{16}$O in the Hartree-Fock (HF) and the
charge- and parity-projected Hartree-Fock (CPPHF) methods as a
function of the momentum squared ($q^2$). The solid, the dashed, and
the dotted curves correspond to the HF calculation, the CPPHF
calculation with the normal tensor force ($x_\text{T}=1.0$), and the
CPPHF calculation with the strong tensor force ($x_\text{T}=1.5$).}
\end{figure}
In Fig.~\ref{fig:o16density} the densities for $^{16}$O in the HF
and CPPHF calculations are shown. Because the tensor force induces
the opposite-parity components with narrower widths in
single-particle wave functions, the densities are depleted in the
middle in the CPPHF calculations compared to the one in the HF
calculation. This effect is larger for the case with the strong
tensor force as expected. To see the effect of the tensor
correlation more clearly, in Fig.~\ref{fig:o16CFF} the charge form
factors are plotted as a function of the momentum squared. From the
figure, higher-momentum components appear in the CPPHF calculation.
It indicates that the tensor force enhances the charge form factor
in a high-momentum region. The short-range correlation, which is not
treated properly here, should have a contribution to the charge form
factor in the high-momentum region. Hence, we need further
investigation to compare the present result of the charge form
factor in the CPPHF method with the experimental data. The
enhancement of the charge form factor in a high-momentum region is
also found in $^4$He in the calculation with the charge- and
parity-projected relativistic mean field model \cite{ogawa06}.
\subsection{Results for the oxygen isotopes}
In this subsection we show the results for the sub-closed-shell
oxygen isotopes.
\begin{figure}
\includegraphics{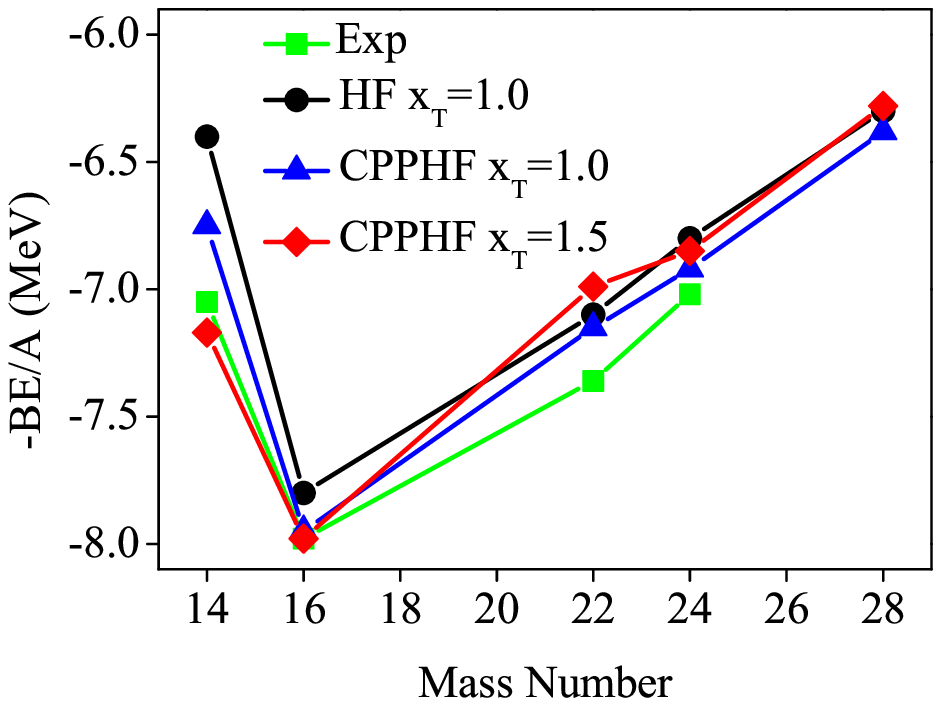}
\caption{\label{fig:BE} (Color online) Binding energies per particle
with minus sign for sub-closed-shell oxygen isotopes. The horizontal
line indicates mass numbers. The circle symbols correspond to the
Hartree-Fock (HF) calculation, the triangle ones to the charge- and
parity projected Hartree-Fock (CPPHF) calculation with the normal
tensor force ($x_\text{T}=1.0$), and the diamond ones to the CPPHF
calculation with the strong tensor force ($x_\text{T}=1.5$). The
square symbols indicate the experimental data \cite{audi03}.}
\end{figure}
In Fig.~\ref{fig:BE}, the results for the binding energies per
particle in the Hartree-Fock (HF) calculation (the circle symbols),
the charge- and parity-projected Hartree-Fock (CPPHF) calculation
with the normal tensor force ($x_\text{T}$=1.0) (the triangle
symbols) and the CPPHF calculation with the strong tensor force
($x_\text{T}$=1.5) (the diamond symbols) are shown. The experimental
data (the square symbols) \cite{audi03} are also plotted.The general
tendency is reproduced in our result, although the agreement with
the experimental data is not as good as the Hartree-Fock-type
calculations \cite{patyk99,bender03,vretenar05,nakada02,nakada06}.
The HF calculation with the MV1 force underestimates the binding
energy of $^{14}$O and $^{22}$O a little bit largely. It indicates
that if we adopt our calculation on the central and the
density-dependent forces in the recent sophisticated effective
interaction, the agreement with the experimental results should
become better. There is an ambiguity in the treatment of the
density-dependent force when we perform the parity and charge
projections, because the density-dependent force cannot be written
in a simple two-body operator form. It causes a difficulty when we
use the density-dependent force in the CPPHF calculation. The
optimization of the central force and the management of the
density-dependent force will be our future problems. The agreement
with the experimental data is good for $^{14}$O in the CPPHF method
with the strong tensor force. In this case quite a large attractive
potential energy comes from the tensor force as shown below.

\begin{figure}
\includegraphics{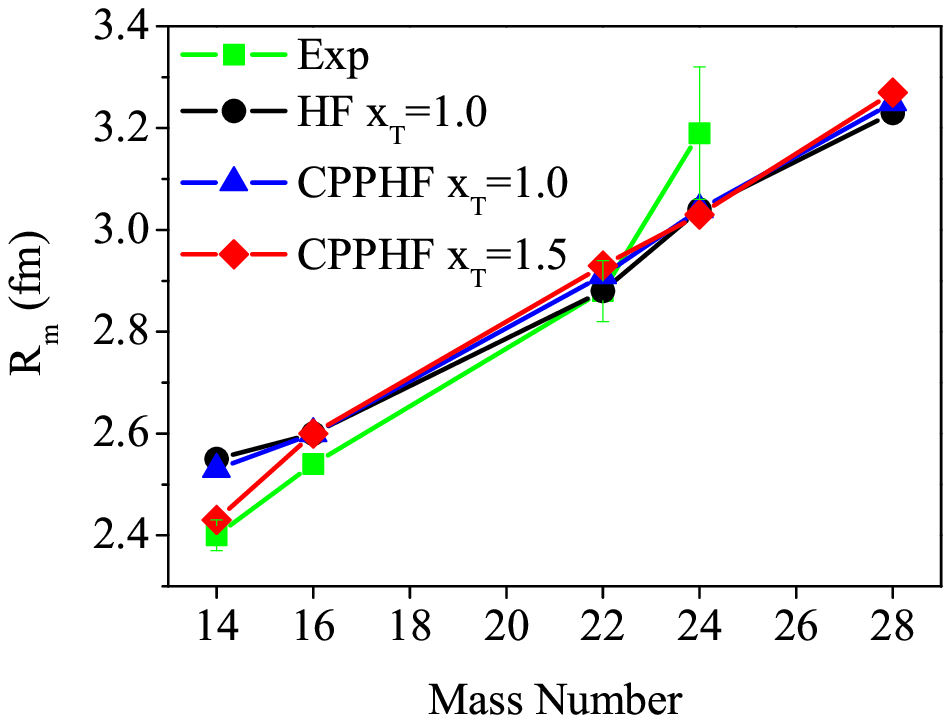}
\caption{\label{fig:Rm} (Color online) Root-mean-square matter radii
for sub-closed-shell oxygen isotopes. The horizontal line indicates
mass numbers. The circle symbols correspond to the Hartree-Fock (HF)
calculation, the triangle ones to the charge- and parity projected
Hartree-Fock (CPPHF) calculation with the normal tensor force
($x_\text{T}=1.0$), and the diamond ones to the CPPHF calculation
with the strong tensor force ($x_\text{T}=1.5$). The square symbols
indicate the experimental data with error bars \cite{ozawa01}.}
\end{figure}
In Fig.~\ref{fig:Rm}, the results for the root-mean-square matter
radii ($R_\text{m}$) are plotted with the experimental data
\cite{ozawa01}. We use the same symbol for each case in
Fig.~\ref{fig:BE}. Except for $^{14}$O the results for all the three
cases are almost the same and reproduce the experimental data well.
For $^{14}$O the CPPHF calculation with the strong tensor force
gives a smaller matter radius compared to the other two calculations
and the agreement with the experimental data becomes better in the
strong tensor force case. The reduction of the radius is caused by
the shrinkage of the opposite parity component in a single-particle
wave function induced by the tensor correlation. Such an effect
cannot be treated in simple Hartree-Fock calculations.

\begin{figure}
\includegraphics{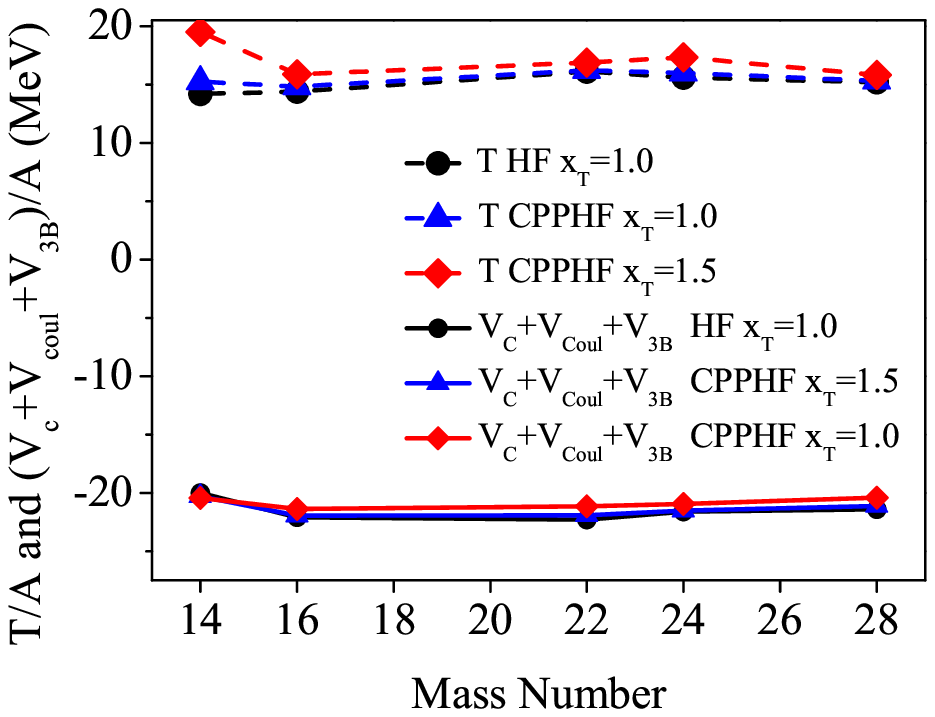}
\caption{\label{fig:TandVC} (Color online) Kinetic energy ($T$) per
particle and the sum of the potential energies from the central
force $V_\text{C}$, the Coulomb force $V_\text{Coul}$, and the
three-body force $V_\text{3B}$ divided by mass numbers. The
horizontal axis indicates mass numbers. The dashed lines correspond
to the kinetic energy and the solid lines to the sum of the
potential energies. The circle, triangle, and diamond symbols
correspond to the Hartree-Fock (HF) calculation, the charge- and
parity-projected Hartree-Fock (CPPHF) calculation with the normal
tensor force ($x_\text{T}=1.0$), and the CPPHF calculation with the
strong tensor force ($x_\text{T}=1.5$).}
\end{figure}
\begin{figure}
\includegraphics{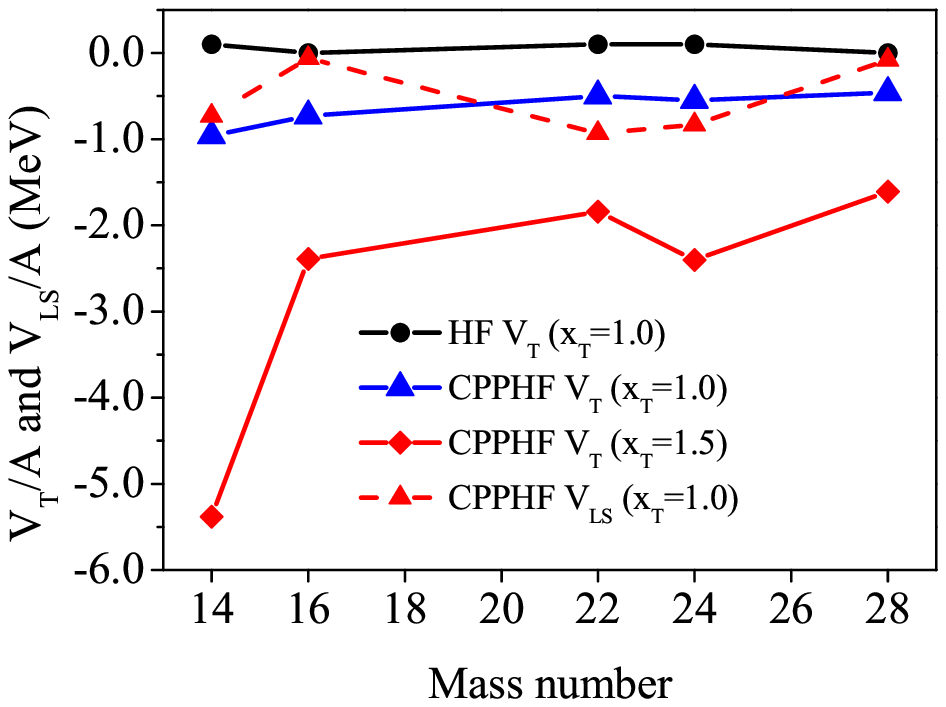}
\caption{\label{fig:VTandVLS} (Color online) Potential energies from
the tensor force ($V_\text{T}$) and the LS force ($V_\text{LS}$)
divided by mass numbers. The horizontal axis indicates mass numbers.
The solid lines correspond to the potential energy from the tensor
force. The dashed line corresponds to the potential energy from the
LS force in the CPPHF calculation with the normal tensor force.
Other results for the LS force are almost unchanged. The meanings of
the symbols are the same as in Fig.~\ref{fig:TandVC}.}
\end{figure}
To see the effect of the tensor force on the binding mechanism in
the oxygen isotopes in Fig.~\ref{fig:TandVC} the results for the
total kinetic energy $T$ (the dashed lines) and the sum of the
potential energies from the central, the three-body and the Coulomb
forces $V_\text{C}+V_\text{3B}+V_\text{Coul}$ (the solid lines) are
plotted. We also show in Fig.~\ref{fig:VTandVLS} the result for the
potential energies from the tensor force $V_\text{T}$ (the solid
lines) and the LS force $V_\text{LS}$ (the dashed line). The same
symbols are used for the HF calculation and the CPPHF calculations
with the normal and strong tensor forces as in Fig.~\ref{fig:BE}.
All values are divided by mass numbers to make a isotope dependence
clear. The total kinetic energy and the sum of the potential
energies from the central, three-body and Coulomb forces show a
volume-like behavior. The kinetic energy for $^{14}$O in the CPPHF
calculation with the strong tensor force is larger than the other
two cases. The increase of the kinetic energy is also caused by the
strong tensor correlation in $^{14}$O, because to gain the
correlation energy from the tensor force the opposite-parity
components must mix into single-particle states and the
opposite-parity components have larger kinetic energy.

The potential energies from the LS force behave in almost the same
manner for the three cases. It becomes attractive if neutron shells
are jj-closed and negligibly small if neutron shells are LS-closed.
The potential energy from the tensor force becomes weakly repulsive
in the HF calculation for all the oxygen isotopes. In the CPPHF
calculations the potential energies from the tensor force become
attractive for all the oxygen isotopes. In contrast to the LS
potential energies, the tensor potential energies have sizable
values in LS-closed shell nuclei like $^{16}$O and $^{28}$O. The
attraction from the tensor force is the same order that from the LS
force even in the case with the normal tensor force. In CPPHF case
$V_\text{T}$ becomes maximum in $^{14}$O and decreases with the mass
number. For the strong tensor force case, the attractive energy from
the tensor force becomes larger as expected. The attractive energy
is quite large for $^{14}$O in the CPPHF calculation with the strong
tensor force. For other isotopes the attractive energies from the
tensor force are small and do not change so much with the mass
number.

\begin{figure}
\includegraphics{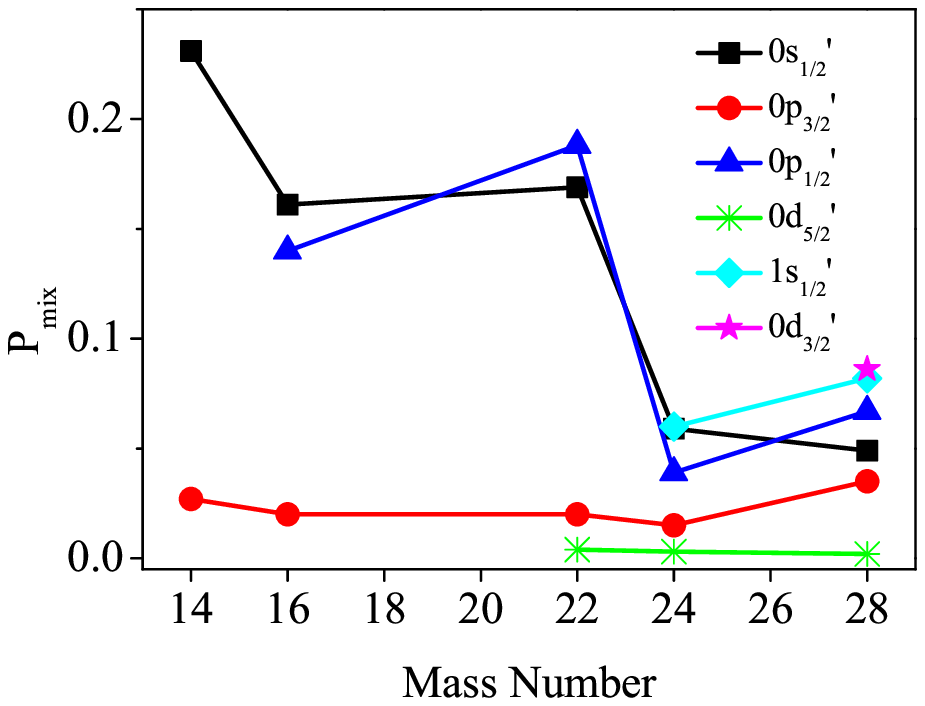}
\caption{\label{fig:PmixNt10} Probabilities of the mixing of the
opposite parity components for single-particle states with neutron
components as dominant ones in the charge- and parity projected
Hartree-Fock calculation with the normal tensor force. The
horizontal axis indicates mass numbers. The square, circle,
triangle, asterisk, diamond, and star symbols correspond to the
first $j$=1/2 ($0s_{1/2}'$; $s_{1/2}$ dominant), the first $j$=3/2
(0$p_{3/2}'$; $p_{3/2}$ dominant), the second $j$=1/2 ($0p_{1/2}'$;
$p_{1/2}$ dominant), the first $j$=5/2 ($0d_{5/2}'$; $d_{5/2}$
dominant), the third $j$=1/2 ($1s_{1/2}'$; $s_{1/2}$ dominant), and
the second $j$=3/2 ($0d_{3/2}'$; $d_{3/2}$ dominant), respectively.}
\end{figure}
In Fig.~\ref{fig:PmixNt10} the probabilities of the mixing of the
opposite parity components $P_\text{mix}$ in the single-particle
states with neutron components as the dominant ones in the result of
the CPPHF calculation with the normal tensor force are shown. The
result for the strong tensor case shows almost the same tendency. In
the usual shell-model classification the first $j$=1/2 state
($s_{1/2}$ dominant), the first $j$=3/2 state ($p_{3/2}$ dominant),
the second $j$=1/2 state ($p_{1/2}$ dominant), the first $j$=5/2
state ($d_{5/2}$ dominant), the third $j$=1/2 state ($s_{1/2}$
dominant), and the second $j$=3/2 state ($d_{3/2}$ dominant)
correspond to $0s_{1/2}$, $0p_{3/2}$, $0p_{1/2}$, $0d_{5/2}$,
$1s_{1/2}$, and $0d_{3/2}$ respectively. All those states are mixed
states of positive and negative parities as in Eq.~(\ref{eq:spwf}).
We add a prime in the following to indicate each single-particle
state has both positive-parity and negative-parity components. For
example, a $0s_{1/2}'$ state has $s_{1/2}$ (positive-parity) and
$p_{1/2}$ (negative parity) components with an $s_{1/2}$ component
as a dominant one. For $^{14}$O the neutron orbits are filled up to
$0p_{3/2}'$. The mixing probability of the opposite parity
components for $0s_{1/2}'$ is larger than 20\% and that for
$0p_{3/2}'$ is a few \% in $^{14}$O. In the CPPHF method the tensor
correlation energy is gained by parity mixing and, therefore,
$P_\text{mix}$ is a measure to indicate how each single-particle
orbit contributes to the tensor correlation. The large
$P_\text{mix}$ for $0s_{1/2}'$ indicates that this orbit is largely
affected by the tensor correlation. In $^{16}$O, the $0p_{1/2}'$
neutron orbit is added newly. The addition of the $0p_{1/2}'$ orbit
reduces the parity mixing of the $0s_{1/2}'$ because they have the
same total angular momentum $j$=1/2. As the result, the tensor
correlation energy becomes smaller. This effect is more significant
for the strong tensor force case as seen in Fig.~\ref{fig:VTandVLS}.
In $^{22}$O the neutron $0d_{5/2}'$ is filled. Because there are no
$j$=5/2 below, $P_\text{mix}$'s for the states filled already in
$^{16}$O do not change largely. In $^{24}$O the $1s_{1/2}'$ orbit is
newly occupied. The occupation of $1s_{1/2}'$ reduces
$P_\text{mix}$'s for the states with $j$=1/2, $0s_{1/2}'$ and
$0p_{1/2}'$. The tensor correlation energy is enhanced, although the
amount of the change is small. Finally, in $^{28}$O the neutron
$0d_{3/2}'$ is filled. $P_\text{mix}$'s for the previously filled
orbits change by this addition but the changes are not so large.

The change of $P_\text{mix}$ as shown above indicates that the large
change of the tensor correlation energy from $^{14}$O to $^{16}$O in
the strong tensor force case is caused by the blocking effect for
the $j$=1/2 orbits. The blocking effect of this kind is shown to be
important in the single-particle $ls$-splitting in $^{5}$He
\cite{myo05}. The effect of the blocking is less significant for the
excess neutron orbits. The effect of the blocking on binding
energies in neutron excess oxygen isotopes seems to be small but the
blocking may affect single-particle natures or collectivity in
neutron excess oxygen isotopes, because the mixing probability is
affected by the blocking effect.

The negligible $P_\text{mix}$ for the $0d_{5/2}'$ orbit indicates
that this orbit does not contribute to the tensor correlation
although there is no other occupied orbits which have $j$=5/2. The
main part of the tensor correlation comes from the $T=0$ channel.
Because the proton $j$=5/2 orbit is not filled in neutron-excess
oxygen isotopes, the $0d_{5/2}'$ orbit is hard to contribute to the
tensor correlation and the mixing probability of it becomes small.

\section{Summary}\label{sec:summary}
We have studied the effect of the tensor force in the
sub-closed-shell oxygen isotopes using the charge- and
parity-projected Hartree-Fock (CPPHF) method. We have extended the
CPPHF method to the cases with Hamiltonians including three-body
forces, although the extension is straightforward. In the CPPHF
method the parity and charge-number projections are performed before
variation. By applying the CPPHF method to the oxygen isotopes
actually, we have found that a sizable potential energy from the
tensor force is obtained in the CPPHF method while in the
Hartree-Fock calculation quite a small potential energy from the
tensor force is obtained.

We have investigated $^{16}$O in some details. The correlation
energy from the tensor force is about 10 MeV for the normal tensor
force case and about 40 MeV for the strong tensor force case. In the
strong tensor force case the strength of the $\tau_1\cdot\tau_2$
channel in the tensor force is multiplied by 1.5. The opposite
parity components induced in single-particle states by the tensor
force have compact sizes as compared to the normal parity
components. This indicates that for the tensor correlation
high-momentum components are important as already found in the alpha
particle case \cite{akaishi04,sugimoto04,myo05,dote05,ogawa06}. The
present result infers that the importance of the high-momentum
component for the tensor correlation is valid in heavier mass
nuclei. We have also shown the density and the charge form factor
calculated in the CPPHF method. The density in the CPPHF method is
reduced around the center and is pulled in to the inside region.
This is caused by the parity mixing and the shrinkage of
single-particle wave functions of opposite parities. The effect of
the shrinkage appears in the charge form factor as a tail in a
high-momentum region, because the shrinkage of the single-particle
wave functions induces high-momentum components in the density.

In the results for the oxygen isotopes, the general tendencies for
the binding energies and the matter radii are reproduced in the
CPPHF calculation with the effective interaction adopted here, while
the agreement of the binding energy with the experimental data is
not so good compared to the Hartree-Fock-type calculations with
recent effective interactions. The root-mean-square matter radii are
well reproduced within error bars except for $^{14}$O with both the
normal tensor force and the strong tensor force. In the strong
tensor force case, the matter radius of $^{14}$O becomes smaller and
close to the experimental data. The reduction of the matter radius
in $^{14}$O with the strong tensor force is due to the shrinkage of
single-particle wave function by the strong tensor correlation,
which is large in $^{14}$O. Because the $0p_{1/2}$ neutron orbit is
not occupied in $^{14}$O, there are no blocking states for the
$0s_{1/2}$ proton orbit in the tensor correlation and, therefore,
the tensor correlation becomes large. Actually, the correlation
energy from the tensor force per particle amounts to more than 5 MeV
in this case. For all the oxygen isotopes the calculated  potential
energy from the tensor force is in the same order as that from the
LS force for the normal tensor force case. In the strong tensor
force case it becomes about two times larger. In contrast to the
potential energy from the LS force, the potential energy from the
tensor force in an LS-closed-shell nucleus does not become close to
zero in the CPPHF calculation. In the Hartree-Fock calculation it
becomes negligibly small because both the total spin and the total
orbital angular momenta is almost zero in an LS-closed-shell
nucleus. The sudden decrease of the potential energy from the tensor
force from $^{14}$O to $^{16}$O is attributed to the blocking effect
of the $j$=1/2 orbits. The blocking effect is also seen in the
neutron-excess oxygen isotopes but does not affect the binding
energy largely.

In the present study we have applied the CPPHF method to the ground
states of the sub-closed-shell oxygen isotopes. The application to
odd-mass nuclei and open-shell nuclei to study the effect of the
tensor force on single-particle natures and the change of
collectivity by the tensor correlation in a neutron excess region
are interesting because the tensor force changes the spin, the
orbital angular momenta and the isospin of nucleon orbits
simultaneously, which is realized in the CPPHF method by the parity
and charge mixing. As for an effective interaction, we combine the
available effective interaction and the tensor force in the free
space with some modifications. We need to use effective interactions
which have the connection with the realistic nuclear forces to
reveal the relation between nuclear structure and the underlying
nuclear force and have the consistency between the tensor force and
other forces like the central and LS forces. The study in such a
direction is also important and now under progress.
\begin{acknowledgments}
We acknowledge fruitful discussions with Prof. H.~Horiuchi on the
role of the tensor force in light nuclei. This work is supported by
the Grant-in-Aid for the 21st Century COE ``Center for Diversity and
Universality in Physics'' from the Ministry of Education, Culture,
Sports, Science and Technology (MEXT) of Japan. This work is
partially performed in the Research Project for the Study of
Unstable Nuclei from Nuclear Cluster Aspects sponsored by the
Institute of Physical and Chemical Research (RIKEN) and a part of
the calculation of the present study was performed on the RIKEN
Super Combined Cluster System (RSCC).
\end{acknowledgments}


\begin{thebibliography}{99}
%
\bibitem{bethe71}
H.A.~Bethe, Annu. Rev. Nucl. Sci. \textbf{21}, 93 (1971).
%
\bibitem{akaishi86}
Y.~Akaishi, in {Cluster Models and Other Topics}, edited by
T.T.S.~Kuo and E.~Osnes (World Scientific, Singapore, 1986), p.~259.
%
\bibitem{kamada01}
H.~Kamada, A.~Nogga, W.~Gl\"{o}ckle, E.~Hiyama, M.~Kamimura,
K.~Varga,
    Y.~Suzuki, M. Viviani, A.~Kievsky, S.~Rosati, J.~Carlson,
    S.C.~Pieper, R.B.~Wiringa, P.~Navr\'{a}til, B.R.~Barrett,
    N.~Barnea, W.~Leidemann, and G.~Orlandini, Phys. Rev. C
\textbf{64}, 044001 (2001).
%
\bibitem{terasawa60}
S.~Takagi, W.~Watari, and M.~Yasuno, Prog. Theor. Phys. \textbf{22},
549 (1959); T.~Terasawa, Prog. Theor. Phys. \textbf{23}, 87 (1960);
A.~Arima and T.~Terasawa, Prog. Theor. Phys. \textbf{23}, 115
(1960).
%
\bibitem{ando81}
K.~And\=o and H.~Band\=o, Prog. Theor. Phys. \textbf{66}, 227
(1980).
%
\bibitem{myo05} T.~Myo, K.~Kat\={o}, and K.~Ikeda, Prog. Theor.
Phys. \textbf{113}, 763 (2005).
%
\bibitem{ozawa00} A. Ozawa, T. Kobayashi, T. Suzuki3, K. Yoshida, and I. Tanihata,
Phys. Rev. Lett. \textbf{84}, 5493 (2000).
%
\bibitem{grawe04}
H. Grawe, Springer Lect. Notes in Phys. \textbf{651}, 33 (2004).
%
\bibitem{otsuka05} T.~Otsuka, T.~Suzuki, R.~Fujimoto, H.~Grawe, and Y.~Akaishi
Phys. Rev. Lett. \textbf{95}, 232502 (2005)
%
\bibitem{toki02} H.~Toki, S.~Sugimoto, and K.~Ikeda,
Prog. Theor. Phys. \textbf{108} 903 (2002).
%
\bibitem{ogawa04} Y.~Ogawa, H.~Toki, S.~Tamenaga, H.~Shen,
A.~Hosaka, S.~Sugimoto, and K.~Ikeda, Prog. Theor. Phys.
\textbf{111}, 75 (2004).
%
\bibitem{sugimoto04} S.~Sugimoto, K.~Ikeda, and H.~Toki, Nucl. Phys.
A \textbf{740}, 77 (2004).
%
\bibitem{ogawa06} Y.~Ogawa, H.~Toki, S.~Tamenaga, S.~Sugimoto, and
K.~Ikeda, Phys. Rev C \textbf{73}, 034301 (2006).
%
\bibitem{sugimoto05} S.~Sugimoto, K.~Ikeda, and H.~Toki,
nucl-th/0511087.
%
\bibitem{akaishi04} Y.~Akaishi, Nucl. Phys. A \textbf{738}, 80
(2004).
%
\bibitem{dote05}
A.~Dot\'{e}, Y.~Kanada-En'yo, H.~Horiuchi, Y.~Akaishi, K.~Ikeda,
Prog. Theor. Phys. \textbf{115}, 1069 (2006).
%
\bibitem{neff03}
T. Neff and H. Feldmeier, Nucl. Phys. A \textbf{713}, 311 (2004).
%
\bibitem{roth06}
R. Roth, P. Papakonstantinou, N. Paar, H. Hergert, T. Neff, and H.
Feldmeier, Phys. Rev. C 73, 044312 (2006).
%
\bibitem{tarbutton68}
R.M.~Tarbutton and K.T.R.~Davies, Nucl Phys. A \textbf{120},1
(1968).
%
\bibitem{bouyssy87}
A.~Bouyssy, J.-F.~Mathiot, N.~Van~Giai, and S.~Marcos, Phys. Rev. C
\textbf{36}, 380 (1987).
%
\bibitem{ando80} T.~Ando, K.~Ikeda, and A.~Tohsaki-Suzuki, Prog.
Theor. Phys. \textbf{64}, 1608 (1980).
%
\bibitem{volkov65}
A.~B.~Volkov, Nucl. Phys. \textbf{74}, 33 (1965).
%
\bibitem{tamagaki68}
R.~Tamagaki, Prog. Theor. Phys. \textbf{39}, 91 (1968).
\bibitem{vautherin72} D.~Vautherin and D.M.~Brink, Phys. Rev. C
\textbf{5} 626 (1972).
%
\bibitem{patyk99}
Z.~Patyk, A.~Baran, J.F.~Berger, J.~Decharg\'e J.~Dobaczewski,
P.~Ring, and A.~Sobiczewski, Phys. Rev. C \textbf{59}, 704 (1999).
%
\bibitem{myo06}
T.~Myo et al. in preparation.
%
\bibitem{CNP1}
P.-G.~Reinhard, in \textit{Cmpulationl Nuclear Phyiscs 1} (edited by
K.~Langanke, J.A.~Maruhn, and S.E.~Koonin, Springer-Verlag, Berlin,
1991) Chapter 2.
%
\bibitem{Flocard73}
H.~Flocard, P.~Quentin, A.K.~Kerman, and D.~Vautherin, Nucl. Phys. A
\textbf{203}, 433 (1973).
%
\bibitem{audi03}
G.~Audi, A. H.~Wapstra and C.~Thibault, Nucl. Phys. A \textbf{729},
337 (2003).
\bibitem{ozawa01}
A.~Ozawa, T.~Suzuki and I.~Tanihata, Nucl. Phys. A \textbf{693}, 32
(2001).
%
\bibitem{bender03} M.~Bender, P.-H.~Heenen, and P.-G.~Reinhard, Rev.
Mod. Phys. \textbf{75}, 121 (2003) and the references therein.
%
\bibitem{vretenar05} D.~Vretenar, A.V.~Afanasjev, G.A.~Lalazissis, and
P.~Ring, Phys. Rep. \textbf{409}, 101 (2005) and the references
therein.
%
\bibitem{nakada02} H.~Nakada and M.~Sato, Nucl. Phys. A \textbf{699},
511 (2002).
%
\bibitem{nakada06} H.~Nakada, Nucl. Phys. A \textbf{764}, 117 (2006).
\end{thebibliography}

\end{document}